\documentclass[floatfix,
 amsmath,amssymb,
 aps,
 prl,
]{revtex4-2}

\DeclareUnicodeCharacter{2212}{-}
\usepackage{multirow}
\usepackage{graphicx}
\usepackage{float}
\usepackage{subcaption}
\usepackage{float}
\usepackage{dcolumn}
\usepackage{bm}

\usepackage[utf8]{inputenc}

\usepackage{siunitx} 
\usepackage{amsmath}
 
\usepackage{xcolor}
\usepackage[utf8]{inputenc}
\usepackage[T1]{fontenc}

\usepackage{amsfonts}
\usepackage{booktabs}

\usepackage{hhline}
\usepackage{array}

\newcolumntype{P}[1]{>{\centering\arraybackslash}p{#1}}

\usepackage{mathtools}

\begin{document}

\title{Experimental Evidence of Cellular Sparing under Single-Pulse Picosecond Irradiation at Extreme Dose Rates}

\newcommand{\QUB}{School of Mathematics and Physics, Queen's University Belfast, BT7 1NN, Belfast, UK}

\newcommand{\NPL}{Radiotherapy and Dosimetry Group, National Physical Laboratory, TW11 0LW, Middlesex, UK}

\newcommand{\PGJCCR}{Johnston Cancer Research Centre, Queen’s University Belfast, BT7 1NN, Belfast, UK}

\newcommand{\qub}{$^1$}
\newcommand{\npl}{$^2$}
\newcommand{\pgjccr}{$^3$}

\author{
H. Maguire\qub,
}
\email{hmaguire11@qub.ac.uk}

\author{
C. A. McAnespie\qub,
P. Chaudhary\npl,
L. Corrigan\qub,
T.~Foster\qub,
S. J. McMahon\pgjccr,
D. Molloy\qub,
G. Nersisyan\qub,
N. O'Connor\qub,
K. M. Prise\pgjccr,
G. Schettino\npl,
~and~
G.~Sarri\qub
}

\address{\qub \QUB} 
\address{\npl \NPL} 
\address{\pgjccr \PGJCCR} 

\begin{abstract}
\section{Abstract}

\textbf{Purpose:} Understanding how biological cells respond to radiation delivered on extremely short timescales is central to resolving the fundamental mechanisms underlying radiotherapy and to defining new regimes of radiobiology. Experimentally accessing these earliest stages of physicochemical response is crucial to benchmark current numerical codes, whose predictions are known to vary substantially with under-constrained early-stage parameters.

\textbf{Methods:} Using a laser-driven electron source, tunable doses up to 8~Gy were delivered to normal (RPE-1 and AG01522-D) and cancer (DU145 and MCF7) human cell lines, in single pulses of picosecond-scale duration at mean dose rates above $10^{10}$~Gy/s. Clonogenic survival assays and 53BP1 foci formation were compared with conventional X-ray irradiation at 2.4~Gy/min.

\textbf{Results:} Surviving fractions are found to be comparable to conventional irradiation at doses up to approximately 3~Gy. At higher doses, however, all cell lines exhibited a consistent increase in surviving fraction, corresponding to an average RBE at 10\% survival of $0.84\pm0.05$. This sparing effect is associated with a statistically significant reduction in the $\beta$ parameter of the linear--quadratic model, indicating reduced cell killing from accumulated sub-lethal damage. Picosecond irradiation also produced a marked reduction in prompt 53BP1 foci, but no significant difference in residual foci at 24~h and foci-size distributions.

\textbf{Conclusions:} These results provide the first experimental evidence of cellular sparing following single-pulse picosecond-scale irradiation. Direct comparison with recent work at femtosecond timescales suggests the presence of a previously undetected sharp transition in biological response at the femtosecond/picosecond interface, giving critical insight on the possible mechanisms responsible for this differential response. The consistency of the response across normal and cancer cell lines suggests a mechanism rooted in early physico-chemical processes rather than cell-line-specific biology, establishing picosecond irradiation as a powerful platform for probing radiobiological response at its earliest stages.
\end{abstract}

\maketitle

\newpage
\section{Introduction}

Radiotherapy (RT) methods are among the most powerful ways to treat cancer, with more than 50\% of cancer patients receiving radiotherapy as part of their treatment worldwide \cite{Pistenmaa_2018,herrera_2017}. Recent developments in radiation delivery and tumour tracking have transformed RT into a precise and personalised treatment. Nonetheless, conventional radiotherapy (CONV-RT) generally suffers from well-known limitations, which are ultimately linked to complications associated with radiation-induced toxicity in normal tissue. These include tissue damage that might require surgical intervention \cite{salvo_2010} and may even result in the development of metastasis and secondary cancers \cite{mahvi_2018}.

A promising technique to differentially target tumour and normal tissue has been identified in FLASH-RT \cite{vozenin_2024}, which delivers high radiation doses in a much shorter time compared to CONV-RT, corresponding to mean dose rates in the region of 10s - 100s of Gy/s and intra-pulse dose rates up to 10$^6$ Gy/s. FLASH-RT has demonstrated a reduction in normal-tissue toxicity while maintaining tumour control comparable to CONV-RT \cite{vozenin_2019,favaudon_2014,vozenin_2024}. Despite this promise, the irradiation conditions required to trigger and maximise the FLASH effect remain uncertain. Dose rate alone is insufficient to define the effect \cite{montaygruel_2020,vozenin_2019}. FLASH-RT must be characterised through several intertwined beam parameters, including total dose, dose per pulse, pulse duration, repetition rate, total irradiation time and radiation quality \cite{Schneider_2022}. This uncertainty reflects a broader lack of consensus on the fundamental mechanisms responsible for FLASH sparing, which remain the subject of intense debate \cite{ma_2024}. Proposed mechanisms include, among others, transient oxygen depletion in the cellular environment \cite{spitz_2019,petersson_2020}, radical--radical recombination \cite{labarbe_2020}, and mitochondrial responses \cite{montaygruel_2020}. Notwithstanding its in vivo evidence, in vitro models have not been able to demonstrate the FLASH effect \cite{vozenin_2024}, with several in vitro studies under FLASH-like irradiation conditions having reported variable and sometimes conflicting results (see, e.g., Refs. \cite{buonanno_2019,adrian_2021,venkatesulu_2019, berry_1969, Ruan_2026}). This variability highlights the sensitivity of the effect to both irradiation parameters and cellular biology. 
In this context, irradiation with effectively \emph{instantaneous} dose delivery could provide a powerful means of isolating the early physico-chemical stages of radiobiological response from later processes, such as oxygen diffusion on millisecond timescales \cite{ma_2024} and DNA damage repair \cite{mcmahon_2018}. 

Laser-driven particle accelerators provide a unique route to this novel regime, as they can generate high-flux particle beams on femtosecond-to-nanosecond timescales with sufficient quality for radiobiological studies  \cite{mcanespie2025single,beyreuther_2015,oppelt_2015,acharya2011dose,
doria2012biological,metzkes2023dresden, obst2026radiation, kroll_2022,giaccaglia_2026,Flacco2025FLASH, MCANESPIE20241105}. 
These irradiation times overlap with the initial physico-chemical stages of cellular response to radiation \cite{adams_1980}, as illustrated in Fig.~\ref{Fig1}. Previous studies investigating biological responses to laser-driven particle beams have spanned different radiation types, dose-deposition timescales, and pulse structures. The findings in each study should thus be interpreted in the context of the specific irradiation parameters, rather than dose-rate alone. For example, the delivery of Gy-scale doses in a single femtosecond-scale electron pulse has been shown to produce a different biological response compared with CONV-RT, including a reduction in cell survival and a higher complexity of damages \cite{mcanespie2025single}.
In contrast, when Gy-scale doses are delivered using \emph{trains} of femtosecond- or picosecond-scale electron pulses (corresponding to mean dose rates in the range of Gy/s-Gy/min) the resulting biological outcomes have been reported to be largely comparable to CONV-RT \cite{giaccaglia_2026,oppelt_2015}. Studies of laser-driven nanosecond-scale proton irradiation with instantaneous dose rates of approximately $10^9$~Gy/s have also reported no observable differences from conventional irradiation for the biological endpoints investigated  (see, e.g., Refs. \cite{doria2012biological,Flacco2025FLASH,hanton2019dna}). Substantial differences in irradiation timescales and linear energy transfer of different irradiation types thus require a high degree of caution in drawing direct comparisons. Nonetheless, 
systematic studies of biological systems to dose deposited on femtosecond-to-picosecond timescales would offer a direct means of probing the earliest stages of radiation response, before the onset of cell-specific biological processes. This genuinely new regime, corresponding to extremely high mean dose rates (EHDR) in the range of $10^9$-$10^{13}$~Gy/s \cite{mcanespie2025single,MCANESPIE20241105,PhysRevE.110.035204}, can thus open a complementary line of investigation that may contribute to a broader understanding of radiation response, including that associated with FLASH-RT.

Here, we report the first comprehensive in vitro study of cellular response to picosecond-scale radiation delivering single-pulse doses tunable up to 8~Gy. Clonogenic survival assays and 53BP1 foci formation and resolution were assessed in two normal cell lines (RPE-1 and AG01522-D) and two cancer cell lines (DU145 and MCF7) grown as monolayers. Results were compared with those obtained using a conventional 160~kVp X-ray source operating at 2.4~Gy/min. For doses up to approximately 3~Gy, no statistically significant difference in surviving fraction is observed between picosecond irradiation and CONV-RT, with a relative biological effectiveness (RBE) at 50\% survival of $1.0\pm0.1$ across all cell lines. At higher doses, however, a consistent increase in surviving fraction following picosecond irradiation is observed, with an RBE at 10\% survival of $0.84\pm0.05$ for all cell lines. Within the linear-quadratic model \cite{mcmahon_2018}, this effect is linked to a statistically significant reduction in the $\beta$ parameter, indicating reduced cell killing from the accumulation of sub-lethal damage. This interpretation is further corroborated by analysis of 53BP1 foci formation and resolution. Remarkably, no statically significant difference in biological response has been observed between the normal and cancer cell lines investigated, suggesting the involvement of fundamental early physico-chemical processes rather than cell-line-specific biology. The findings reported here thus hint at at a response characteristic of picosecond-scale radiation, providing a line of investigation complementary to FLASH-RT.

These results reveal a distinct biological response that also differs fundamentally from multi-pulse picosecond irradiations \cite{giaccaglia_2026} and from femtosecond-scale irradiation \cite{mcanespie2025single}.  The observed reduction in the $\beta$ parameter is consistent with observations at femtosecond timescales \cite{mcanespie2025single}. However, femtosecond-scale dose delivery has also been associated with a significant increase in the $\alpha$ parameter, commonly linked to prompt single-track lethal damage, leading to an overall reduction in surviving fractions \cite{mcanespie2025single}. That increase in $\alpha$ is not observed here. Together, these findings thus reveal a previously unobserved transition in cellular response at the femtosecond--picosecond interface, which is not captured by numerical models \cite{perl2012topas,agostinelli2003geant4}. Further, these results establish picosecond-scale single-pulse irradiation as a novel and robust platform for investigating cellular response at timescales comparable to the earliest physico-chemical stages of radiation response. This is a critical tool for benchmarking current radiobiology numerical codes \cite{perl2012topas,agostinelli2003geant4},  which are known to exhibit large variations in predicted outcomes as a function of under-constrained early-stage parameters, such as direct damage threshold, chemical stage length and hydroxyl-radical damage probability \cite{Zhu_2020}.

\section{Methods and Materials} 

\subsection{Electron Beam Source}

The experiment was performed using the TARANIS laser facility at Queen's University Belfast \cite{dzelzainis_2010}. The laser delivered an energy of E$_L$ = (10.6 $\pm$ 0.5)~J in a pulse duration of $\tau_L$ = (0.8 $\pm$ 0.1)~ps and at a central wavelength $\lambda$ = 1.053~$\mu $m. Uncertainties correspond to the standard deviation in laser beam parameters across the experimental campaign. An F/3 off-axis parabola was used to focus the laser onto a 20~$\mu$m gold foil at an incidence angle of 30°. This produced a focal spot with a full-width at half maximum of $(5.4 \pm 0.1)$ $\mu$m $\times$ $(6.5 \pm 0.2)$ $\mu$m containing $(33.2 \pm 0.2 )\%$ of the total laser energy, resulting in a peak intensity on target of (2.1 $\pm$ 0.2)$\times 10 ^{19}$~W/cm$^2$.

\begin{figure*}[b!]
\centering
 \includegraphics[width=0.9\linewidth]{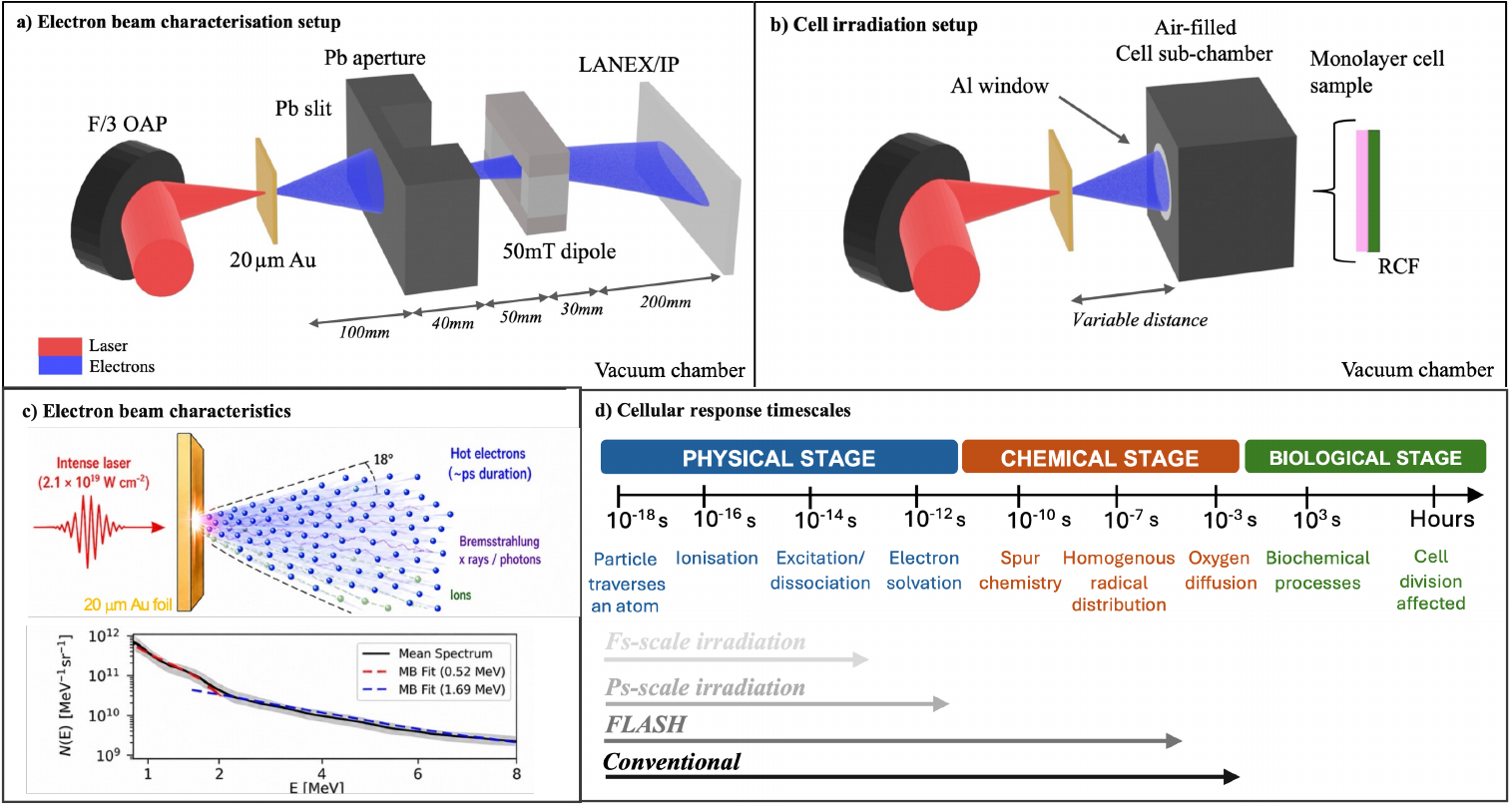}

\caption{
Experimental setups for \textbf{(a)} the characterisation of the laser-driven electron source and \textbf{(b)} in vitro cell irradiation. \textbf{(c)} Illustration of the electron beam generation mechanism and measured averaged spectrum of accelerated electrons (grey shading represents the standard deviation of two shots) whereas blue and red dashed lines show a two-temperature Maxwellian fit. \textbf{(d)} Comparison between indicative timescales of cellular response to radiation and timescales associated with different irradiation modalities (fs-scale $\simeq$ 100~fs, ps-scale $\simeq$ 10~ps, FLASH $\simeq$ $\mu s$ - ms, Conventional $\simeq$ s - minutes.)}
\label{Fig1}
\end{figure*}

The interaction of the focussed high-intensity laser with the solid target produces a high-charge electron beam with multi-MeV energy \cite{macchi_2013} (see Fig. \ref{Fig1}). To first characterise the electron beam (Fig. \ref{Fig1}.a), a lead collimator and a magnetic spectrometer were positioned after the solid target.  A 1~mm wide lead slit was placed 102~mm from the gold foil (angular acceptance of 10 mrad), followed by a 40~mm thick, 10~mm diameter lead (Pb) aperture acting as a secondary collimator.  A permanent dipole magnet was positioned 50~mm from the Pb aperture; mapping
of the magnetic field distribution inside the dipole showed a super-Gaussian (index=4) magnetic field with $\sigma = 15$~mm and peak magnetic field $B_{max} = 53$~mT \cite{Audet_2021}. Deflected electrons were recorded using either image plates (IP) and LANEX scintillator screens mounted 200~mm downstream of the rear of the dipole. 

The electron beam exhibited a Maxwellian-like energy spectrum, with a two-temperature distribution (T$_{cold}$ = (0.52 $\pm$ 0.05)~MeV and T$_{hot}$ = (1.7 $\pm$ 0.1)~MeV, see Fig. \ref{Fig1}.c) and a measured divergence of 36.0$^{\circ}$ $\pm$ 4.6$^{\circ}$ and 37.2$^{\circ}$ $\pm$ 3.1$^{\circ}$ in the two transverse directions, consistent with the $\vec{J} \times \vec{B}$ heating mechanism \cite{jxb,green_2008}. The duration of the electron beam at source is estimated to be approximately 1.3 times the laser pulse duration \cite{2006NatPh...2...48F} (of the order of 1 picosecond) and the total number of electrons per shot was measured to be on the order of $10^{11}$, with typical values around $5.5\times 10^{11}$.


\subsection{Irradiation Conditions}
For cell irradiation, the collimators and the magnetic dipole were removed from the beam path and replaced with a sub-chamber as shown in Fig. \ref{Fig1}.b. The sub-chamber comprised a 1~mm-thick aluminium window facing the beam, with the cell samples secured 18~mm downstream and an RCF positioned directly behind each sample. The sub-chamber was sealed under ambient and sterile conditions maintaining the cells at atmospheric pressure and isolated from the vacuum of the main laser chamber.
Due to the large divergence of the electron beam, the delivered dose could be controlled by varying the propagation distance in vacuum from the source of the electron beam to the sub-chamber aluminium window, while maintaining all the other interaction conditions. 
In-vacuum distances from the source to the aluminium window were varied between 15 and 57~mm. As the cell layer was positioned 18~mm inside the sub-chamber, the corresponding source-to-cell distances ranged from 33 to 75~mm. 

To determine whether varying the source-to-cell distance altered the radiation quality to the cells, Monte Carlo simulations using the TOPAS code version 3.9 \cite{perl2012topas, agostinelli2003geant4} were performed to assess the electron energy spectrum at the cell plane. The simulations fully reproduced the experimental irradiation geometry (see Supplementary Material S1 for details) and confirmed no appreciable change in the spectral shape over the experimental range of source-to-cell distances (33 - 75~mm). Furthermore, electron LET in water varies only within 10\% over the relevant energy range of 0.5 to 8 MeV \cite{NIST_ESTAR}. This confirms that the reduction in delivered dose with increasing source-to-cell distance is associated only with the reduction in electron fluence resulting from beam divergence, rather than changes in electron energy spectrum or LET (see Supplementary Material S1 for details \cite{suppl}). 

A conventional 160 kVp X-ray source (CONV; MultiRad160, Precision X-ray, Inc., Madison, CT, USA) operating at a dose rate of 2.4 Gy/min, was used as the reference radiation. The mean LET at the cell irradiation plane was approximately 1.75 keV/µm for the X-ray field, compared with approximately 0.18–0.20 keV/µm for electrons across the energy range relevant to the present study. Despite this difference, both radiation qualities are within the low-LET regime and previous experimental comparisons have demonstrated comparable clonogenic effectiveness between kilovoltage X-rays and MeV electrons \cite{spadinger1992relative, wanstall2024first}, as also supported by numerical modelling \cite{Delorme2021} (further details in Supplementary Material S1 \cite{suppl}).

 \subsection{Cell Cultures and Analysis}

Four cell lines (DU145, MCF7, RPE-1 and AG01522-D) were cultured under standard conditions in their recommended media. Biological response was assesed using clonogenic cell survival assays and DNA double-strand break damages.

For EHDR clonogenic cell survival irradiations, cells were seeded onto customised stainless-steel dishes fitted with a 3~$\mu$m Mylar membrane and housed in an air-filled sub-chamber positioned within the vacuum chamber. The sub-chamber was mounted on a motorised stage that rotated the dishes into a vertical orientation immediately before irradiation. This ensured that the electron beam did not interact with the medium, preserving longitudinal dose uniformity. Following irradiation, dishes were returned to a horizontal position. For CONV irradiation, cells were irradiated in standard six-well plates using the MultiRad 160 system. For both modalities, corresponding unirradiated controls were removed from the incubator for the duration of the irradiation procedure and maintained alongside the respective experimental set-up to account for the differing handling conditions and determine plating efficiency (PE). Sham controls were also performed to assess the potential effect of the EHDR irradiation procedure, with no significant differences observed between sham-treated and control samples. Further details are provided in Supplementary Material S2 \cite{suppl}.

Clonogenic survival was determined 12 days post-irradiation and surviving fraction calculated as:
\begin{equation}
SF = \frac{\text{Colonies}}{\text{Cells Plated} \times PE}
\end{equation}
Relative biological effectiveness (RBE) was calculated as the ratio of the CONV to EHDR dose required to achieve the same surviving fraction, with RBE$_{50}$ and RBE$_{10}$ determined at surviving fractions of 50\% and 10\%, respectively.

\vspace{1mm}

DNA double-strand break repair kinetics were quantified by measuring the persistence of radiation induced foci at 0.5, 1, 6 and 24 h following irradiation. A minimum of $2\times10{^5}$ cells were plated in slide flasks (Thermo Scientific™ Nunc™ Lab-Tek™ Flask on Slide) and allowed to adhere overnight. The same irradiation vessels were used for both EHDR and CONV exposures, with samples irradiated to $\approx$ 1~Gy. Detailed cell culture, clonogenic assay, immunofluorescence staining and image analysis procedures are provided in Supplementary Material S2 \cite{suppl}.

\subsection{Dosimetry}
Gafchromic™ EBT3 radiochromic films (RCFs) were used to measure the on-shot dose for each individual biological irradiation from the picosecond electron beam. 
RCF calibration was performed using the CONV 160~kVp X-ray source, with doses ranging from 1 - 10~Gy (see Supplementary Material S3 \cite{suppl} and Ref. \cite{PhysRevE.110.035204}. for additional details).
EBT films have demonstrated limited dose-rate dependence and good agreement with independent dosimetric approaches under a range of pulsed irradiation conditions, including picosecond electron pulses and doses per pulse comparable to those used in the present study \cite{karsch_doserate_indep, Bazalova_Carter_doserate_indep, beyreuther2010establishment, jaccard2017high}. These studies support the use of EBT3 film without a dose-rate-dependent correction for the present measurements. More recently, a dose-dependent over-response of EBT3 films at ultra-high dose rates was reported with statistical significance emerging above 3~Gy \cite{Horst_2026}. As shown in Supplementary Material S3 \cite{suppl}, the potential influence of this effect was assessed by applying the dose-dependent correction reported by Horst et al. and re-analysing the clonogenic survival data. As discussed in the following, the application of such correction did not alter the main conclusions of this study.

\subsection{Dose Properties at the Cell Irradiation Plane}

\begin{figure*}[b!]
\centering
\includegraphics[width=0.95\linewidth]{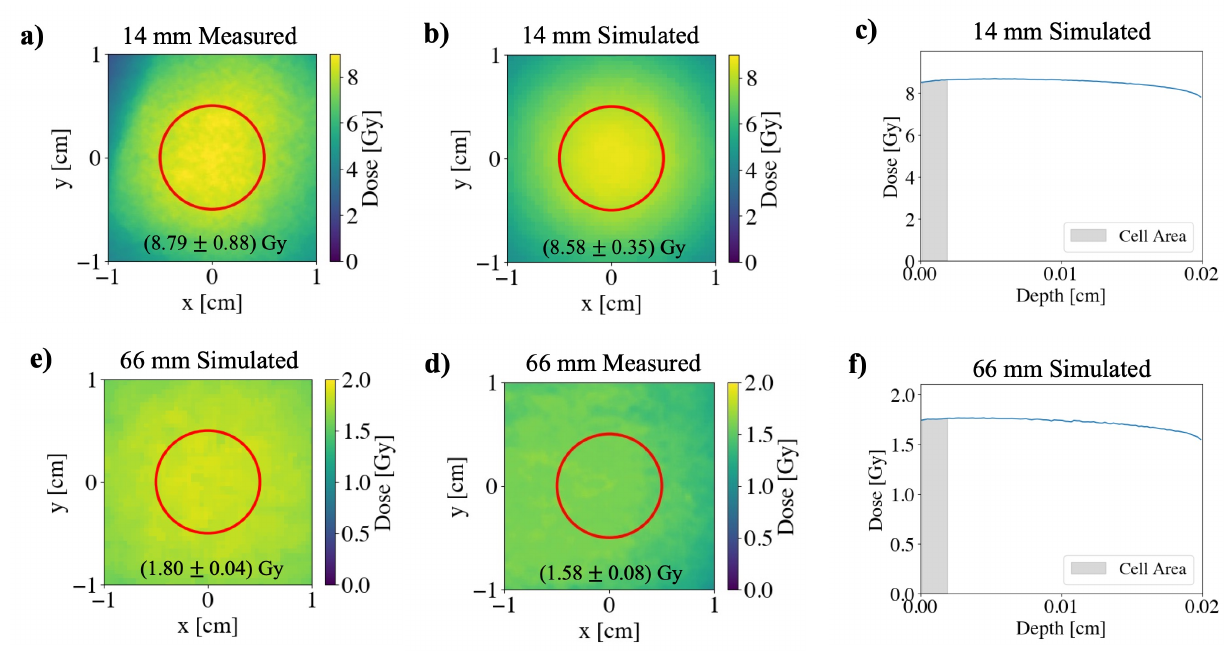}
\vspace{1mm}
\caption{Experimental dose profiles and results from matching TOPAS simulations 14~mm and 66~mm from the source. \textbf{(a, b)} Measured and simulated dose profiles at 14~mm. \textbf{(d, e)} Measured and simulated dose profiles at 66~mm. The red circle indicates the cell sample position (1~cm diameter). \textbf{(c, f)} Simulated dose-depth profiles at 14 mm and 66 mm, illustrating a high degree of longitudinal dose uniformity in the cell volume (grey-shaded areas).}
\label{fig:RCF}
\end{figure*}

Dose characterisation was performed prior to cell irradiation using the 1~mm aluminium window and RCFs in a geometry reproducing the cell irradiation set-up. The aluminium window and the RCFs were translated together along the beam axis to vary their distance from the electron source while maintaining their relative positions.
Figs. \ref{fig:RCF}.a and \ref{fig:RCF}.d show representative examples of the dose deposited by the laser-accelerated electron beam 14~mm and 66~mm from the source, respectively. The red circles indicate the positions of the cell samples, each with a 1~cm cross-sectional diameter. 
In these examples, the measured doses were $D_{14mm} = 8.79\pm0.88$~Gy and $D_{66mm} = 1.58\pm0.08$~Gy. The uncertainty corresponds to the transverse variation across the irradiated area, with coefficients of variation (CoV) of the order of 10\% and 5\%, respectively. Unlike the RCF-only dose characterisation measurements, the cell irradiation geometry was constrained by the dimensions of the cell sub-chamber, such that cell samples could not be positioned closer than 33~mm from the electron source. This corresponded to a maximum accessible dose for cell irradiation of $7.56\pm0.67$~Gy.

Dose profiles and dose-depth distributions were also obtained from matching TOPAS simulations with examples shown in Fig. \ref{fig:RCF}. 
Across the central 1~cm diameter region, the simulated doses were $D_\mathrm{14~mm} =  ($8.58 $\pm$ 0.35)~Gy and $D_\mathrm{66~mm} =  ($1.80 $\pm$ 0.04)~Gy, in good agreement with the experimental measurements. The corresponding dose-depth profiles are shown in Fig. \ref{fig:RCF}.c and Fig. \ref{fig:RCF}.f. The adoption of cell monolayers (shaded area in the figures) allowed for a flat dose-depth profile across the cell sample (longitudinal CoV of 0.48\% and 0.30\% for the two examples), thus minimising uncertainties in deposited dose.

\begin{figure}[b!]
\centering
\hspace*{-0.5cm} 
\includegraphics[width=0.95\linewidth]{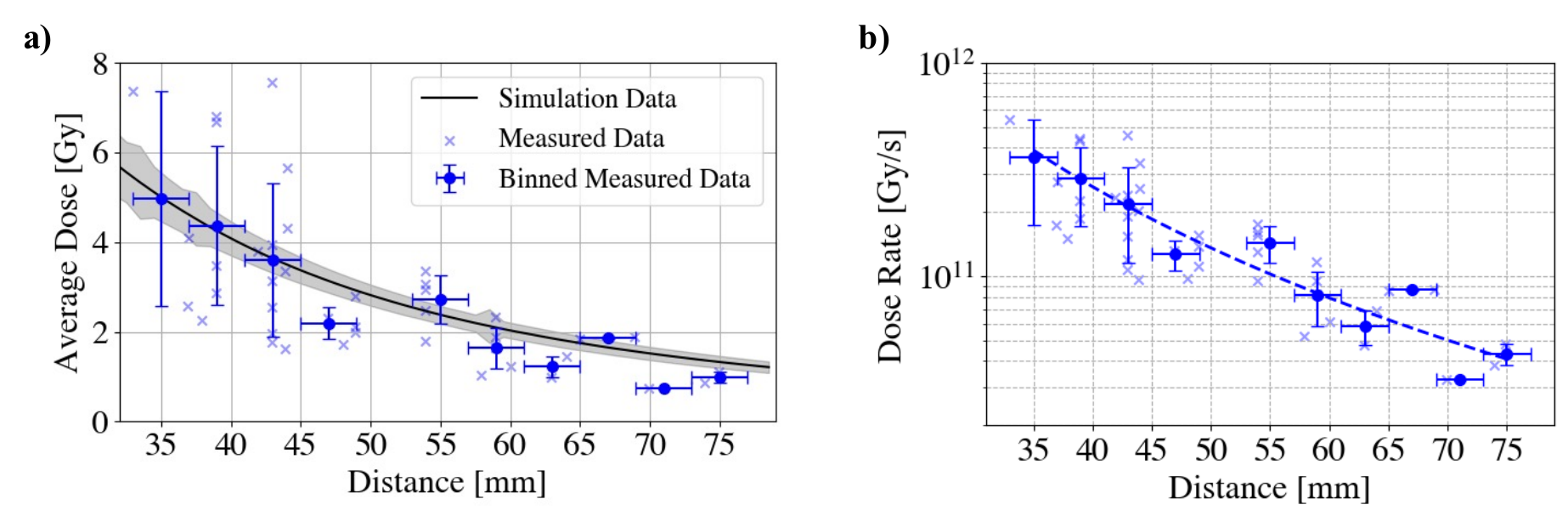}
\vspace{1mm}
\caption{\textbf{(a)} Measured dose (blue crosses) deposited across the irradiation region as a function of distance for 40 experimental shots (blue dots from the binning in 4~mm increments), compared with corresponding simulated doses (black curve). The grey shaded region represents the statistical uncertainty in the simulated dose, obtained from five independent simulation runs at each position. \textbf{(b)} Dose rate as a function of distance of the cell sample from the electron beam source for the same shots. The dashed line shows a $1/d^3$ fit to the binned data.}
\label{fig:dose_v_distance}
\end{figure}

The experimentally measured doses at each position over 40 shots are shown in Fig. \ref{fig:dose_v_distance} together with the corresponding Monte-Carlo simulations. Good agreement was observed between measurements and simulations, with the deposited dose decreasing as a function of distance $d$ following a $1/d^2$ dependence. Fluctuations in dose between separate irradiations are predominantly due to fluctuations in the laser-solid interaction conditions. Importantly, this shot-to-shot variability did not introduce uncertainty to the results reported as on-shot dose measurements were obtained for each irradiation.

The electron beam has a duration at source of the order of 1.3 times the laser beam duration \cite{2006NatPh...2...48F}, corresponding to approximately 1~ps. However, the broad energy spectrum of the beam implies that, during free propagation to the cell samples, the beam experiences a longitudinal spread due to the different time of flight of each energy slice. As a result, the electron beam has a 1/$e^2$ duration of approximately 12.5~ps 33~mm away from the target and 23~ps at 75~mm. Consequently, the mean dose rates at different distances from the target varied between $3$$\times$$10^{10}$ and $5$$\times$$10^{11}$~Gy/s (see Fig. \ref{fig:dose_v_distance}.b). The dose rate decreased with distance following a $1/d^3$ dependence (Fig. \ref{fig:dose_v_distance}.b), with a quadratic term corresponding to the geometrical decrease in dose and an additional linear term from the lengthening of the electron beam due to time-of-flight effects.

The interaction of a high-intensity laser with a solid target can also generate dense population of ions via Target Normal Sheath Acceleration \cite{macchi_2013} and X-rays, which can in principle contribute to the dose deposited. In similar conditions, the TARANIS laser produces broadband proton beams with a maximum cut-off energy of 12~MeV \cite{dzelzainis_2010}, which have a range in aluminium of approximately 0.87~mm \cite{NIST_PSTAR}.  Therefore, the protons co-travelling with the electron beam are effectively stopped by the 1~mm Al window in front of the cell monolayers. Laser-accelerated protons are thus not expected to contribute significantly to the dose deposited on the cell samples or the RCFs, as also confirmed by matching simulations. X-rays are also generated in the target and during bremsstrahlung of the electrons as they propagate through the 1~mm Al window. Further simulations were run to estimate the contribution of X-rays to the dose observed at the cell plane. Simulations show that the X-ray beam at the cell irradiation plane has a fluence approximately four orders of magnitude lower than the electron beam. At the closest and furthest experimental distances of 33~mm and 75~mm to the cell sample, the total doses deposited by X-rays were 2.5~mGy and 0.2~mGy, respectively. In both cases, and all other intermediate positions, the total dose deposited by X-rays accounts for less than 1\% of the total dose and can thus be neglected.



\vspace{1mm}

\section{Results} 

\subsection{Clonogenic Cell Survival}

\sisetup{
  detect-all,          
  separate-uncertainty = true, 
  table-align-uncertainty = true,
  table-number-alignment = center
}

\begin{figure}[b!]
\centering

\includegraphics[width=0.9\linewidth]{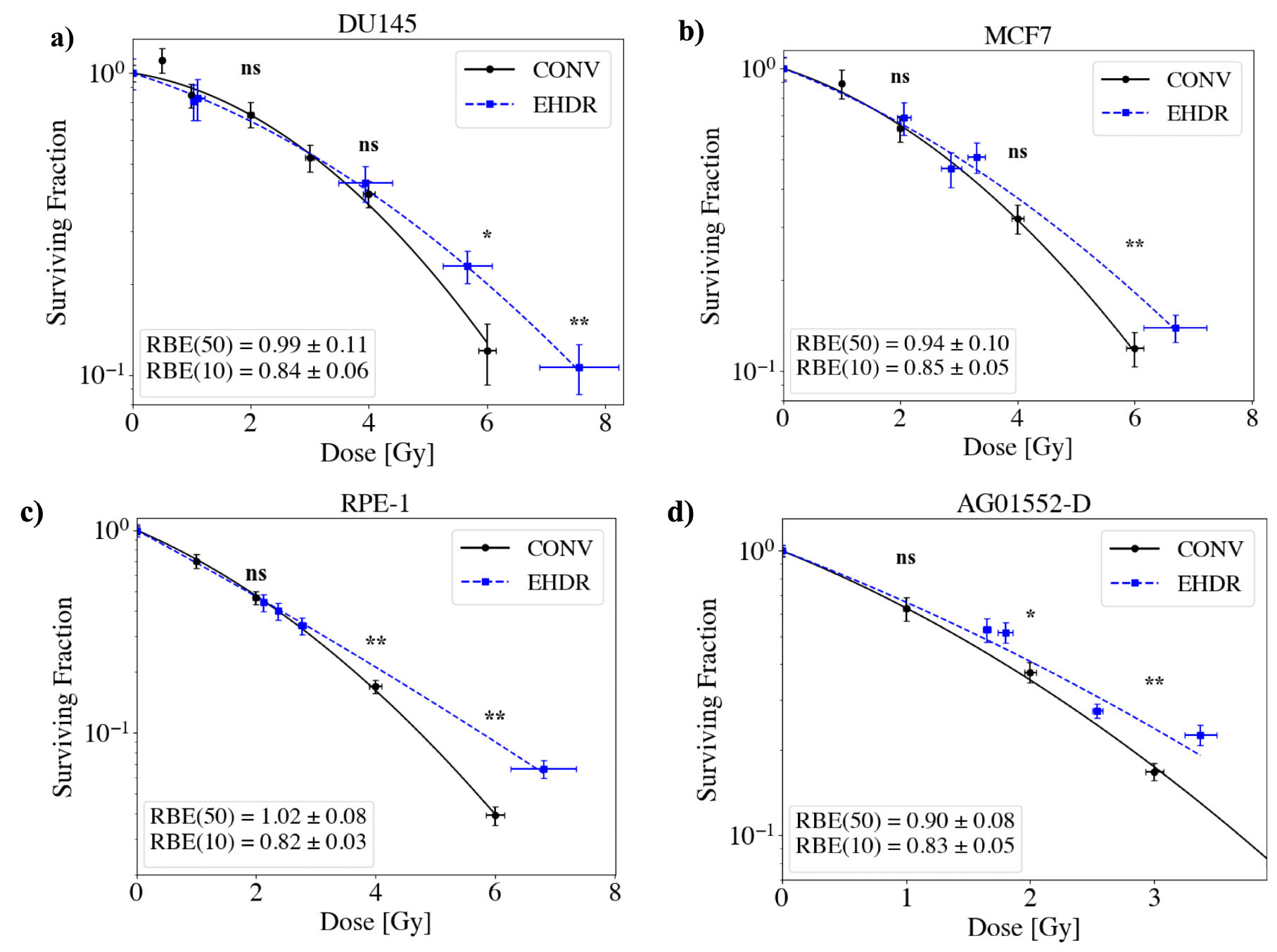}
\vskip 10mm

\setlength{\tabcolsep}{10pt}
\renewcommand{\arraystretch}{1.3}
\resizebox{\textwidth}{!}{%
\begin{tabular}{
l
l l l
l l l
c c c
}
\hline
\textbf{Cell Line }
& \multicolumn{3}{c}{\textbf{CONV}} 
& \multicolumn{3}{c}{\textbf{EHDR}} 
& \multicolumn{1}{c}{\textbf{RBE (50)}} 
& \multicolumn{1}{c}{\textbf{RBE (10)}} 
& \multicolumn{1}{c}{\textbf{F-test}} \\
\cline{2-7}
 & \multicolumn{1}{c}{$\alpha$} 
 & \multicolumn{1}{c}{$\beta$} 
 & \multicolumn{1}{c}{$\alpha/\beta$}
 & \multicolumn{1}{c}{$\alpha$} 
 & \multicolumn{1}{c}{$\beta$} 
 & \multicolumn{1}{c}{$\alpha/\beta$}
 & \multicolumn{1}{c}{} 
 & \multicolumn{1}{c}{} 
 & \multicolumn{1}{c}{\textit{p}-value} \\
\hline
\textbf{a) DU145}     
& 0.07 $\pm$ 0.04 & 0.05 $\pm$ 0.01 & 1.4 $\pm$ 0.8 
& 0.14 $\pm$ 0.06 & 0.02 $\pm$ 0.01 & 7.0 $\pm$ 4.6 
& 0.99 $\pm$ 0.11& 0.84 $\pm$ 0.06& 0.01\\

\textbf{b) MCF7}      
& 0.14 $\pm$ 0.05 & 0.04 $\pm$ 0.01 & 3.5 $\pm$ 1.5 
& 0.17 $\pm$ 0.05 & 0.02 $\pm$ 0.01 & 8.5 $\pm$ 4.9 
& 0.94 $\pm$ 0.10 & 0.85 $\pm$ 0.05 &  0.00\\

\textbf{c) RPE-1}     
& 0.29 $\pm$ 0.04 & 0.04 $\pm$ 0.01 & 7.3 $\pm$ 2.1 
& 0.36 $\pm$ 0.03 & 0.01 $\pm$ 0.00 & 36.0 $\pm$ 4.7 
& 1.02 $\pm$ 0.08 & 0.82 $\pm$ 0.03 & 0.00 \\

\textbf{d) AG01522-D}& 0.40 $\pm$ 0.07 & 0.06 $\pm$ 0.02 & 6.7 $\pm$ 2.5 
& 0.38 $\pm$ 0.06 & 0.03 $\pm$ 0.02 & 12.7 $\pm$ 8.7 
& 0.90 $\pm$ 0.08 & 0.83 $\pm 0.05^\dagger$& 0.05 \\
\hline
\end{tabular}
}

 \vspace{1mm}
\caption{Surviving fraction assessed by clonogenic assays comparing picosecond-scale electron (EHDR) irradiation (blue) and a conventional X-ray source (black) for human in vitro cell lines: \textbf{(a)} DU145, \textbf{(b)} MCF7, \textbf{(c)} RPE-1 and \textbf{(d)} AG01522-D. The solid lines illustrate the fitted survival curve according to the linear quadratic model (Eq. 1). RBE at 50\% and 10\% survival are presented on each plot. Statistical significance was evaluated using two-tailed Z-tests at regular dose intervals: 2, 4, 6 \& 7.56 ~Gy for DU145, 2, 4 \& 6 ~Gy for MCF7 and RPE-1 and 1, 2 \& 3 ~Gy for AG01522-D.  Results are denoted as: \textbf{ns} not significant, \textbf{*} p < 0.05, \textbf{**} p < 0.01.
The table summarises the corresponding $\alpha$ and $\beta$ values, RBE values and F-test \textit{p}-values. $^\dag$ The RBE(10) value for the AG01522-D cell line is extrapolated from the fitting parameters.
}
\label{fig:SF}
\end{figure}

Clonogenic surviving fractions as a function of dose delivered by the picosecond-scale electron beams (EHDR) for four cell lines (DU145, MCF7, RPE-1 and AG01522-D) are shown in Fig. \ref{fig:SF} and compared with the CONV X-ray source. CONV irradiations were performed in triplicates, with 6 technical replicates per independent experiment. Data points are presented as the mean and standard deviation across the three experiments. EHDR measurements were performed within a single experimental campaign, presented as the mean and standard deviation of 6 technical replicates per irradiation. All datasets are fitted using the linear-quadratic model
\begin{equation}
SF(D) = e^{-(\alpha D  + \beta D^2)},
\end{equation}
where $\alpha$ and $\beta$ are fitting parameters qualitatively representing the relative weight of lethal DNA damage from single hits and accumulation of sub-lethal damages, respectively \cite{mcmahon_2018}. Data was fitted using a non-linear regression method weighted by standard deviation. 
This initial fit was followed by Markov Chain Monte Carlo (MCMC) sampling to estimate robust posterior distributions for the model parameters. 
For each cell line, the data was fitted both as a unified dataset and independently for the two irradiation modalities; the F-test was then used to determine whether the fit was significantly improved by using separate parameters. A summary of the main fitting parameters and the corresponding $p-$values from the F-test are shown in Fig. \ref{fig:SF}. For all cell lines, the F-test provides strong evidence against the null hypothesis, confirming that the surviving curves are statistically different for the two irradiation modalities. 

Motivated by the results of the global F-test, dose-specific differences in surviving fraction were subsequently evaluated. As identical dose values could not always be achieved experimentally between CONV and EHDR irradiations, the linear-quadratic models were re-parametrised to directly estimate surviving fraction and uncertainty at predefined doses. The LQ fits were weighted by the experimental uncertainties in surviving fraction and the uncertainty in the model-estimated surviving fraction was obtained from the posterior distributions of the fitted $\alpha$ and $\beta$ parameters. Comparisons were performed at 2, 4, 6 and 7.56~Gy for DU145, 2, 4 and 6~Gy for MCF7 and RPE-1 cells, and at 1, 2 and 3~Gy for AG01522-D cells.
Differences between the two irradiation modalities were assessed using a two-tailed Z-test (statistical significance overplotted in Fig. \ref{fig:SF}). This analysis enabled identification of the doses at which statistically significant differences in surviving fraction were observed. 

For all cell lines, the surviving fractions show good agreement between the CONV and EHDR irradiation modalities up to doses of the order of 2-3 Gy. This is reflected by an RBE at 50\% survival of the order of 1 for all cell lines. However, a statistically significant difference between the two irradiation modalities is consistently observed for all cell lines at higher doses. This is reflected by an RBE at 10\% survival of the order of 0.84 for all cell lines, a result comparable with FLASH-like sparing effects reported in the literature \cite{Wilson_2020, vozenin_2019,adrian_2021}. The uncertainties in RBE were calculated by propagating the uncertainty from the MCMC-derived parameter distributions. In the case of AG01522-D, reduced laser performance prevented clonogenic assay results from being obtained for doses significantly above 3~Gy. However, extrapolation of the survival curves at higher doses indicate a behaviour consistent with the other cell lines.   
Another two tumour cell lines, HeLa S3 and H460, were also irradiated with single-pulse doses up to 2 and 3~Gy, respectively, with the surviving fractions provided in Supplementary Material S4 \cite{suppl}. Also for these cell lines, no significant difference between CONV and EHDR irradiation was observed for single-pulse doses up to 3 Gy, consistently with the results reported in Fig. \ref{fig:SF}.


A potential non-linearity in the response of EBT3 films to high doses per pulse has recently been reported \cite{Horst_2026}, with an over-response of up to 7\% reported for doses of the order of 8~Gy. This potential correction was applied to our experimental results (see Supplementary Material S3 \cite{suppl}) and did not alter the main conclusions of the study, with the overall differences between EHDR and CONV survival curves remaining statistically significant ($p\leq0.05$, F-test) for all four cell lines.

The observed sparing effect is consistently associated with a significant increase in the $\alpha/\beta$ ratio across all cell lines, driven primarily by a pronounced reduction in the $\beta$ parameter. This may reflect decreased lethal damage arising from the accumulation of multiple sub-lethal lesions and is discussed further in the Discussion section.

\subsection{DNA Double-Strand Break Damage}

\begin{figure}[ht]
\centering
\includegraphics[width=0.90\linewidth]{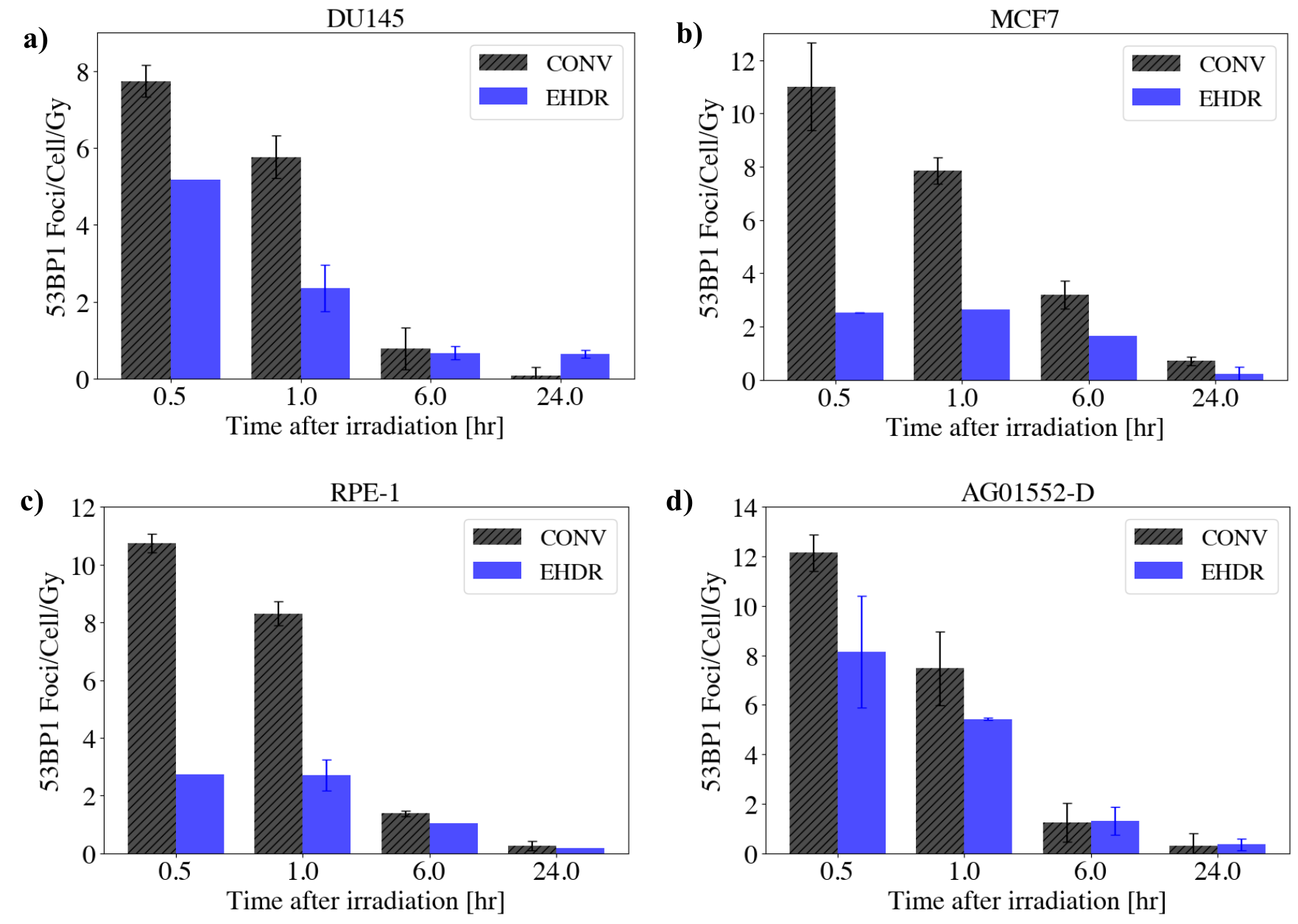}
\vspace{1mm}
\caption{Radiation-induced DNA-double strand break foci using 53BP1 in \textbf{(a)} DU145, \textbf{(b)} MCF7, \textbf{(c) }RPE-1 and \textbf{(d)} AG01522-D cell lines. Blue bars represent the EHDR response while black bars represent the CONV response.}
\label{fig:foci}
\end{figure}

\begin{figure*}
\centering
\includegraphics[width=0.9\linewidth]{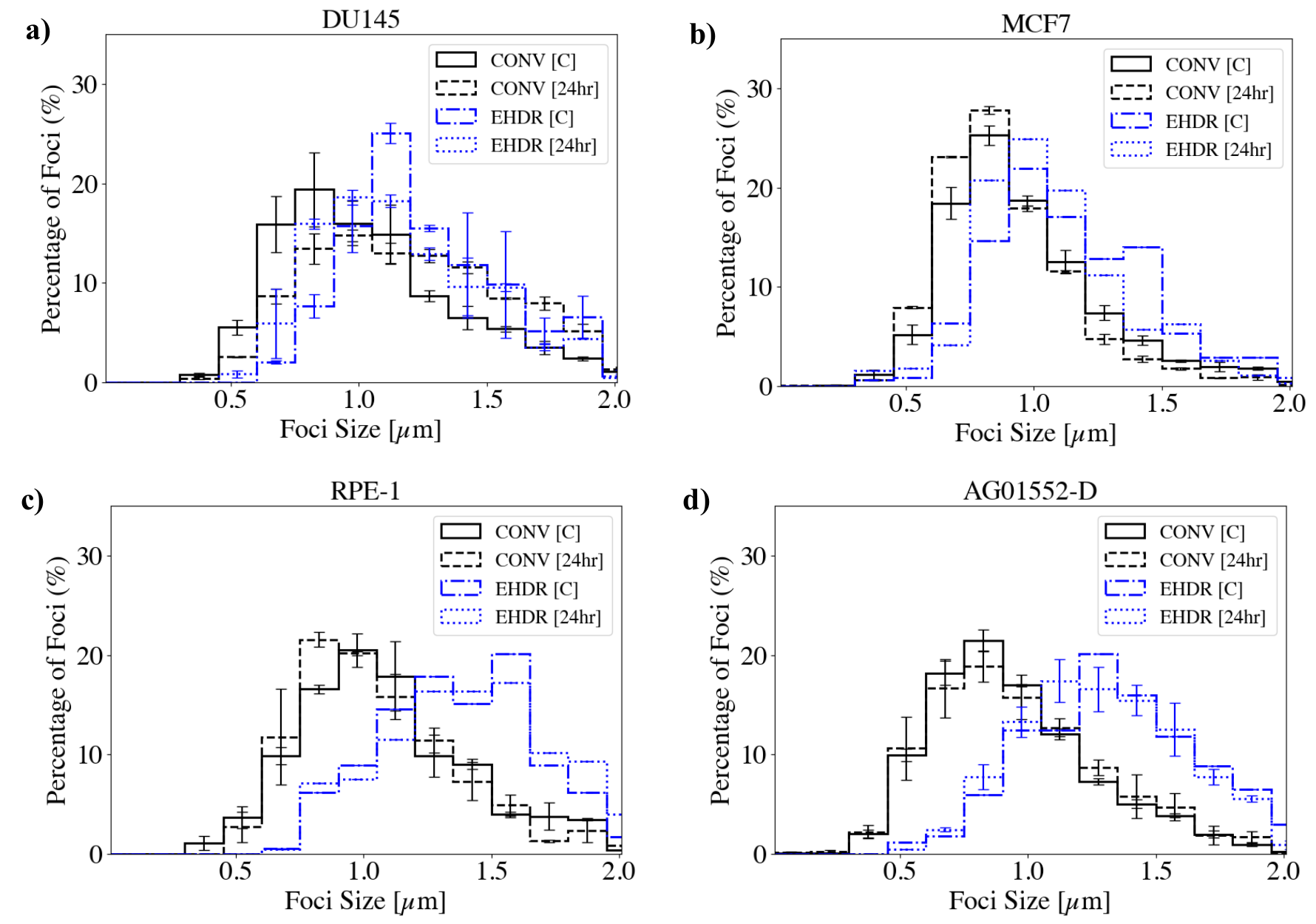}
\vskip 10mm 

\centering
\setlength{\tabcolsep}{10pt}
\renewcommand{\arraystretch}{1.2}
\makebox[\textwidth][c]{%
\begin{tabular}{l l c c c c}

\toprule
\textbf{Cell line} & \textbf{Source} & \multicolumn{2}{c}{\textbf{Foci size / $\mu$m}} & \textbf{\% Diff} & \textbf{\textit{p}-value} \\
\cmidrule(lr){3-4}
 & & \textbf{Control} & \textbf{24~hour}& & \\
\midrule
DU145     & CONV & 1.04 $\pm$ 0.35 & 1.20 $\pm$ 0.37 & 15.31 & 0.00 \\
          & EHDR & 1.24 $\pm$0.30 & 1.16 $\pm$ 0.32 & -6.12 & 0.00 \\
MCF7      & CONV & 0.96 $\pm$ 0.31 & 0.89 $\pm$ 0.26 & -7.19 & 0.00 \\
          & EHDR & 1.14 $\pm$ 0.31 & 1.07 $\pm$ 0.29 & -6.19 & 0.00 \\
RPE-1     & CONV & 1.08 $\pm$ 0.33 & 1.05 $\pm$ 0.31 & -3.25 & 0.17 \\
          & EHDR & 1.36 $\pm$ 0.30 & 1.40 $\pm$ 0.32 & 3.10 & 0.47 \\
AG01522-D & CONV & 0.94 $\pm$ 0.32 & 0.95 $\pm$ 0.33 & 1.09 & 0.22 \\
          & EHDR & 1.33 $\pm$ 0.32 & 1.29 $\pm$ 0.31 & -3.27 & 0.09 \\
\bottomrule
\end{tabular}}

\vspace{2mm}

\caption{Histograms of the size of DNA-double strand break foci using 53BP1 from control (unirradiated samples, dashed lines) and the 24~hour time point (solid lines) in \textbf{(a)} DU145, \textbf{(b)} MCF7, \textbf{(c) }RPE-1 and \textbf{(d)} AG01522-D cell lines. Blue and black lines represent the EHDR and CONV response, respectively. The table summarises the mean $\pm$ standard deviation of the size distributions. The percentage difference (\% Diff) is calculated with respect to the mean of the control sizes and the \textit{p}-values from Kolmogorv-Smirnov tests are reported. }
\label{fig:foci size}
\end{figure*}

Radiation-induced DNA double-strand breaks (DSBs) were quantified by 53BP1 foci formation and resolution at 0.5, 1, 6 and 24~hours post-irradiation. Results are shown in Figs. \ref{fig:foci} and \ref{fig:foci size}. CONV data represent the mean ± SD of three independent repeats. EHDR data represent one or two independent repeats depending on the experimental condition, with error bars indicating the SD where two independent repeats were obtained. No error bars are shown for conditions for which only one independent measurement was obtained.

\vspace{1mm}
All cell lines exhibit comparable behaviour, with a significantly lower number of prompt DNA damages (0.5 - 1 hour after irradiation) for EHDR. However, the number of foci persisting 24 hours after irradiation is not observed to be significantly different between CONV and EHDR irradiation for all cell lines. Even though a general reduction of prompt DNA damage has been previously observed for FLASH-like in vitro irradiations (see, e.g., Ref. \cite{fouillade_2020}), the reduction appears to be significantly more marked in this irradiation regime, an observation that warrants further investigation.

To further characterise the DNA DSBs, the foci sizes at the 24~hour time point (measured as the full width at half maximum, FWHM) were also analysed (see Fig. \ref{fig:foci size}). For each radiation source, the distribution of foci sizes at 24~hours post irradiation was compared with that obtained from control slides. Percentage differences in mean size were calculated with respect to the control distributions and statistical differences were assessed using the Kolmogorov–Smirnov test. Normal cell lines (RPE-1 and AG01522-D) do not show a significant difference between control and irradiated samples in both irradiation regimes, whereas a small but statistically significant difference is observed for the tumour cell lines (DU145 and MCF7). However, no statistically significant difference is observed across irradiation regimes for each cell line. Even though it must be noted that the distribution of foci size for low-LET irradiation can only yield qualitative information \cite{roobol_2020}, this observation potentially indicates similar repair patterns \cite{francisco2008induction, 10.1158/0008-5472.CAN-04-1601}, in contrast with the increase in foci size observed for single-pulse femtosecond-scale irradiation \cite{mcanespie2025single}.

\section{Discussion}

We have reported on the first irradiation platform able to deliver multi-Gy doses in a single picosecond-scale pulse with sufficient quality to perform radiobiological studies. 
Single-pulse doses tunable up to 8~Gy were achieved, allowing for a comprehensive characterisation of cellular response at 
mean dose rates in the range of $3\times10^{10}-5\times10^{11}$~Gy/s. The use of cell monolayers and the adoption of key improvements in dose delivery and control allowed for a clear observation of different biological outcomes in this regime when compared to irradiation at a conventional dose rate of 2.4~Gy/min and recently reported results using femtosecond-scale deliveries at mean dose rates above $10^{13}$~Gy/s \cite{mcanespie2025single}.

Surviving fractions closely agree with X-ray irradiation at 2.4~Gy/min for doses up to 2-3~Gy, as elucidated by an average RBE at 50\% survival of $1.0\pm0.1$ for all cell lines. However, all cell lines showed a clear departure in surviving fractions at higher doses, with an average relative biological effectiveness at 10\% survival of $0.84\pm0.05$. 

The sparing effect observed in this study is associated with a consistent and significant decrease of the $\beta$ parameter in the linear-quadratic model, a strong indication of a reduction of cell killing induced by accumulation of sub-lethal damages \cite{mcmahon_2018}. The surviving fractions for the EHDR and CONV modalities are effectively indistinguishable at low doses. This strongly suggests that the $\alpha$ parameter in the linear-quadratic model, the dominant factor at low doses, is similar for both modalities. To test this hypothesis, the surviving fractions for EHDR were refitted with $\beta$ as the only free parameter, while $\alpha$ was fixed to the value obtained from fitting the CONV surviving fractions (see Supplementary Material S5 \cite{suppl}). These fits evidence a consistent and significant reduction in the quadratic component of the linear-quadratic model for all cell lines, with 
$\beta_{\text{CONV}}/\beta_{\text{EHDR}}$ of the order of 2 
($1.4\pm0.3$, $1.5\pm0.5$, $2.3\pm0.5$ and $2.4\pm1.0$ for DU145, MCF7, RPE-1 and AG01522-D, respectively). 

Clonogenic assays are complemented by DNA double-strand break studies through 53BP1 foci formation. Foci evolution over 24~hours after irradiation show a general reduction of prompt foci (0.5 to 1~hour after irradiation) but a comparable number of foci after 24~hours, when compared to a conventional dose rate. This is qualitatively consistent with in vitro studies of FLASH irradiation at dose rates in the range of 10s -100s~Gy/s (see, e.g., Refs. \cite{fouillade_2020, adrian_2021}), even though a significantly more marked reduction of prompt DNA damages is observed in this irradiation regime. Early foci reflect the initial detection of DNA lesions but do not distinguish between repairable and lethal clusters. The larger reduction in prompt foci thus suggests that picosecond-scale processes may alter the onset of early chemical processes. In contrast, the convergence of residual foci at 24 hours is consistent with similar repair patterns, indicating that the yield of complex, unrepairable damage remains largely unchanged across irradiation modalities. This is consistent with an unchanged $\alpha$ parameter in the linear-quadratic model of surviving fractions.

Additionally, no significant differences in 53BP1 foci sizes were observed for each cell line across the two irradiation regimes. For low-LET radiation, focus size is not necessarily a direct surrogate for DNA damage complexity but instead might reflect changes in repair dynamics \cite{roobol_2020}, further corroborating the hypothesis of similar repair patterns following picosecond-scale irradiation. 

The observed sparing effect is not expected to be linked with oxygen diffusion or metabolic oxygen consumption, both of which are negligible on picosecond timescales \cite{adams_1980}. However, the duration of dose delivery overlaps with the earliest stages of water radiolysis and radical chemistry, raising the possibility of a transient, highly localised oxygen depletion driven by radiochemical processes \cite{Alanazi_2021}. In this picture, the rapid production of reducing species during an intense single pulse could transiently lower the local oxygen concentration on nanometre length scales, thereby reducing the fixation of radiation-induced DNA lesions even under nominally normoxic conditions \cite{pratx_2019}. Since radiolytic oxygen consumption is approximately proportional to the number of reducing species produced, it is expected to scale linearly with dose, unless local oxygen exhaustion or strong nonlinear feedback occurs \cite{Scarmelotto_2024}. A reduction in cell killing driven by radiolytic oxygen consumption would thus manifest itself as a net reduction in the quadratic term of the surviving fractions. This effect would be independent of cell-specific biology or repair mechanisms, in agreement with the consistency of this effect across substantially different cell lines.

Interestingly, a comparable reduction in the $\beta$ parameter has been recently observed also for single-pulse irradiation at femtosecond timescales \cite{mcanespie2025single}. However, at those timescales, this was accompanied by a higher $\alpha$ parameter (RBE at 50\% =$1.4\pm0.1$) and a statistically significant increase in foci size, suggesting additional onset of more complex prompt damages. This may indicate a similar reduction in sub-lethal damages across the femtosecond-to-picosecond interface, but with a higher likelihood of prompt lethal damages at femtosecond timescales. These results suggest the presence of a sharp transition in cellular behaviour between femtosecond and picosecond timescales, the underlying causes of which are, to the best of our knowledge, currently unknown.

The observation that femtosecond- and picosecond-scale irradiation produce distinct biological outcomes may have important implications for mechanistic modelling. This is particularly relevant because current Monte Carlo codes, including Geant4-DNA and TOPAS-nBio \cite{perl2012topas,agostinelli2003geant4}, show significant variability in predicted outcomes depending on under-constrained early-stage parameters, such as direct damage threshold, chemical stage duration and hydroxyl-radical damage probability \cite{Zhu_2020}. 


Although extensive and coordinated experimental and modelling efforts will be required to fully unravel cellular responses in this regime, the present work establishes a robust platform for probing previously inaccessible irradiation conditions on timescales directly overlapping the earliest physico-chemical stages of radiation response and enabling benchmarking and extensions of such models. The first evidence that distinct biological effects can emerge across the femtosecond–picosecond transition also opens new lines of investigation into the response of biological cells to ultrashort, extremely high dose rate radiation.\\



\bibliography{Refs}
\clearpage
\end{document}